# High-Fidelity In Silico Dataset of Ordered Binary and Ternary MXenes with Predictive Synthesis Guidance

Ahmed H. Biby[1] and Charles B. Musgrave* [2,3,4]
[1]Materials Science and Engineering Program, University of Colorado Boulder, Boulder, CO 80309, USA
[2]Department of Chemical and Biological Engineering, University of Colorado, Boulder, Colorado 80303, USA
[3]Department of Chemical Engineering, University of Utah, Salt Lake City, Utah, 84112, USA
[4]Department of Materials Science and Engineering, University of Utah, Salt Lake City, Utah, 84112, USA

## Abstract

Despite their promise, the targeted synthesis of novel MXenes remains challenging due to their vast structural and compositional space and the lack of reliable stability descriptors that account for synthesis conditions. To address this gap, we utilized density functional theory (DFT) with the $r^2$SCAN functional to develop a high-fidelity dataset of 1,040 ordered binary and ternary bare delaminated MXenes, including mono-transition metal MXenes, in-plane ordered double-transition metal MXenes, in-plane ordered metal-divacancy MXenes, and out-of-plane ordered double-transition-metal MXenes. For each structure, we calculated formation and decomposition Gibbs energies, decomposition products, layer symmetry groups, and energies above the hull. We find that all MXenes are thermodynamically inaccessible under equilibrium conditions. However, 115 predominantly carbide MXenes become thermodynamically stabilized at elevated temperatures (500–2000 K), and 23 are stabilized under high pressure (10 GPa). Additionally, three MXenes in the Mn–Fe–C chemical space are stabilized within an experimentally feasible range of carbon chemical potentials. These findings reveal stability trends across the MXene space and identify synthesis strategies that can guide experimental efforts. Our dataset provides a robust foundation for developing machine-learned stability descriptors and for guiding experimental efforts toward novel MXene discovery. By identifying synthesizable candidates along with their synthesis conditions, this work bridges a critical gap in data-driven approaches to MXene discovery and lays the foundation for future autonomous experimental workflows.

## Introduction

MXenes are layered nanomaterials derived from layered MAX phases that exhibit tunable composition, hydrophilicity, metallic conductivity, and mechanical robustness.[1] MXenes have the general formula $M_{n+1}X_nT_z$, where M is an early transition metal, X is typically either C or N, T represents a surface termination group resulting from the synthesis process, and n typically ranges from 1 to 3.[2] X atoms reside in the octahedral interstitial sites, while M forms either a hexagonal close-packed (hcp) or face-centered cubic (fcc) lattice at n = 1 and n > 1, respectively.[3] The MXene family encompasses a wide range of ordered structural space and is commonly classified according to metal-site ordering. Following established MXene nomenclature, these include mono-transition metal MXenes (m-MXenes, $M_{n+1}X_nT_z$), where n ranges from 1 to 3; in-plane ordered double-transition metal MXenes (i-MXenes, $M_{4/3}M'_{2/3}XT_z$); in-plane ordered metal-divacancy MXenes (v-MXenes, $M_{4/3}\square_{2/3}XT_z$); and out-of-plane ordered double-transition-metal MXenes (o-MXenes, $M_{n-1}M'_2X_nT_z$), where n ranges between 2 and 3, as illustrated in **Figure** 1.[4] MXenes also exhibit disordered structures, such as solid solutions and high-entropy structures.[4,5] These diverse compositional and structural spaces lead to tunable bandgaps, electrical conductivity, photoelectronic properties, and mechanical properties.[6] Consequently, MXenes are ideal for a plethora of applications such as catalysis, energy conversion, energy storage, biomaterials, sensors, and electromagnetic interference shielding.[2,7] Although MXenes possess an extensive compositional and structural design space, to date fewer than 50 have been experimentally synthesized and fewer than 100 theoretically investigated,[2,3,8–10] representing only a small fraction of the theoretical MXene chemical space. The main challenge in synthesizing novel MXenes is the reliance on heuristics and chemical intuition. Unfortunately, no empirical or machine-learned descriptor currently exists that reliably predicts the likelihood that a composition forms a stable MXene structure, rendering it challenging to evaluate the vast multinary chemical space.

MXene properties can be accurately predicted using quantum mechanical methods, such as density functional theory (DFT). To the best of our knowledge, the aNANT open-access online repository is the only MXene-dedicated dataset available.[11] This dataset comprises single-layer (delaminated) MXenes with the general formula MM'XTT', where the transition metals (M and M') and surface terminations (T and T') are either the same or different. aNANT encompasses 11 possible metal elements for M, 2 possible X elements (C or N), and 14 possible terminating groups, with these combinations yielding a total of 23,870 MXenes. Despite the large number of entries, the dataset is limited to o-MXene structures with either two identical metallic layers or dissimilar metallic layers that have not yet been experimentally reported. Furthermore, aNANT does not report thermodynamic stability data, and its structures were optimized using the low-fidelity PBE DFT exchange-correlation functional. On the other hand, the Materials Project repository (MP) contains only four multi-layer (laminated) MXenes: mp-1094034 ($Ti_3C_2$), mp-1246153 ($Ti_3N_2$), mp-1094023 ($V_2C$), and mp-1096885 ($Cr_3N_2$).[12] The Computational 2D Materials Database (C2DB)[13,14] is a well-established repository that reports structural, thermodynamic, elastic, electronic, magnetic, and optical properties of 2D materials across more than 30 distinct crystal structures. Among its 16,905 entries, only 72 are MXenes, all of which are delaminated m-MXenes. The limited volume, variety, and fidelity of MXene data hinder the computational exploration of MXenes synthesizability. This emphasizes the need for systematic and consistent computational studies aimed at developing comprehensive MXene datasets. Bridging this gap is essential for enabling machine learning to accelerate the design, discovery, and synthesis of novel MXenes.

In this work, we employed the regularized strongly constrained and appropriately normed density functional ($r^2$SCAN), which is known to offer improved accuracy over PBE for predicting ground-state structures and formation energies of solids.[15] As an initial step towards the systematic development of a MXene dataset aligned with FAIR ("Findable, Accessible, Interoperable, and Reusable") principles, we deposited the dataset in Zenodo, as described in the Data Records section. We used the $r^2$SCAN functional to generate a high-fidelity dataset comprising 1,040 ordered binary and ternary bare (no terminating groups) delaminated MXene carbides and nitrides.

For each DFT-converged structure, we report the formation Gibbs energies ($\Delta G_f$), decomposition Gibbs energies ($\Delta G_d$), and energies above the hull ($E_{hull}$) relative to elemental references and competing phases, as well as decomposition products, and layer symmetry groups. In addition, we investigated experimentally-feasible constrained equilibrium by tuning thermodynamic constraints (temperature, pressure, and chemical potential) to assess the accessibility of metastable

MXenes.[16,17] Here, “constrained equilibrium” refers to conditions at which the system effectively equilibrates within a restricted chemical environment because formation of the global equilibrium phases is kinetically inhibited, enabling metastable phases to persist.[16] Thus, we studied the effects of temperature, pressure, and chemical potential on the stability of MXenes and identified MXenes that can be stabilized along with their minimum onset synthesis conditions. This dataset enables the identification of potentially undiscovered MXenes via a predictive synthesis approach, supports the development of structure–composition–stability trends through analyzing each studied MXene with respect to its stability, and provides a foundation for training machine-learned MXene stability descriptors.

All of the MXenes analyzed herein were found to be thermodynamically inaccessible; that is, they lie above the convex hull under ambient equilibrium conditions. However, 115 mostly m- and o-MXene carbides become stable at elevated temperatures (500–2000 K), and 23 are stabilized under 10 GPa of applied pressure. Additionally, some Mn–Fe–C ternary MXenes are stabilized within an experimentally feasible carbon chemical potential range of –6 to 1 eV/C atom. These findings highlight the viability of accessing new MXenes through thermodynamically constrained, non-equilibrium synthesis routes. This work reports a high-fidelity computational MXene dataset aimed at aiding the development of autonomous labs to explore MXene synthesizability and the thermodynamic synthesis conditions required to enable targeted synthesis for advanced technological applications.

## Methodology

## Structure generation and data scope.

We retrieved the MAX phases $Ti_2AlC$ (mp-12990), $Ti_3AlC_2$ (mp-3747), and $Ti_4AlN_3$ (mp-568934) from MP.[12] We created single-layer (delaminated) MXenes by removing the Al atoms that connect layers. We then used topotactic structure templating to generate initial MXene structures by elemental substitutions on parent MXene lattices. The `DLSVolumePredictor`,[18] which uses data-mined lattice scaling was then used to obtain better initial structures while maintaining at least 15 Å of vacuum slab to avoid spurious effects of periodic images. With C and N as the two possible X-site occupants, we investigated M-site substitutions using 13 early transition metals, representing the broadest range of d-block elements reported in the MAX/MXene literature. M includes Sc, Ti, V, Cr, Mn, Fe, Y, Zr, Nb, Mo, Hf, Ta, and W.[19] Combinatorial permutations of these M and X elements yielded 78 m-MXenes ($M_{n+1}X_n$, where n ∈ {1, 2, 3}), 312 i-MXenes ($M_{4/3}M'_{2/3}X$), 624 o-MXenes ($M_{n-1}M'_2X_n$, where n ∈ {2, 3}), and 26 v-MXenes ($M_{4/3}\square_{2/3}X$). To ensure clarity of MXene structural nomenclature and avoid ambiguities, such as $M_4X_3$ referring to either v- or m-MXenes, we used a naming convention tailored to the scope of this study, as illustrated in **Figure 1**. It consists of two parts separated by a dash. The first part is a letter indicating the MXene category, as previously described in the introduction, and the second part is the value of n in the MXene category formula, as explained above. For binary (m- and v-) MXenes, the second part represents the M and X elements, respectively, while for ternary (o- and i-) MXenes, it represents the M, M’, and X elements, respectively. In the case of o-MXenes, M indicates the inner metal layer(s), while M’ represents the outer metal layers. For i-MXenes, M is the element that occupies two-thirds of the metal sites, and M’ the element that occupies the remaining one-third.

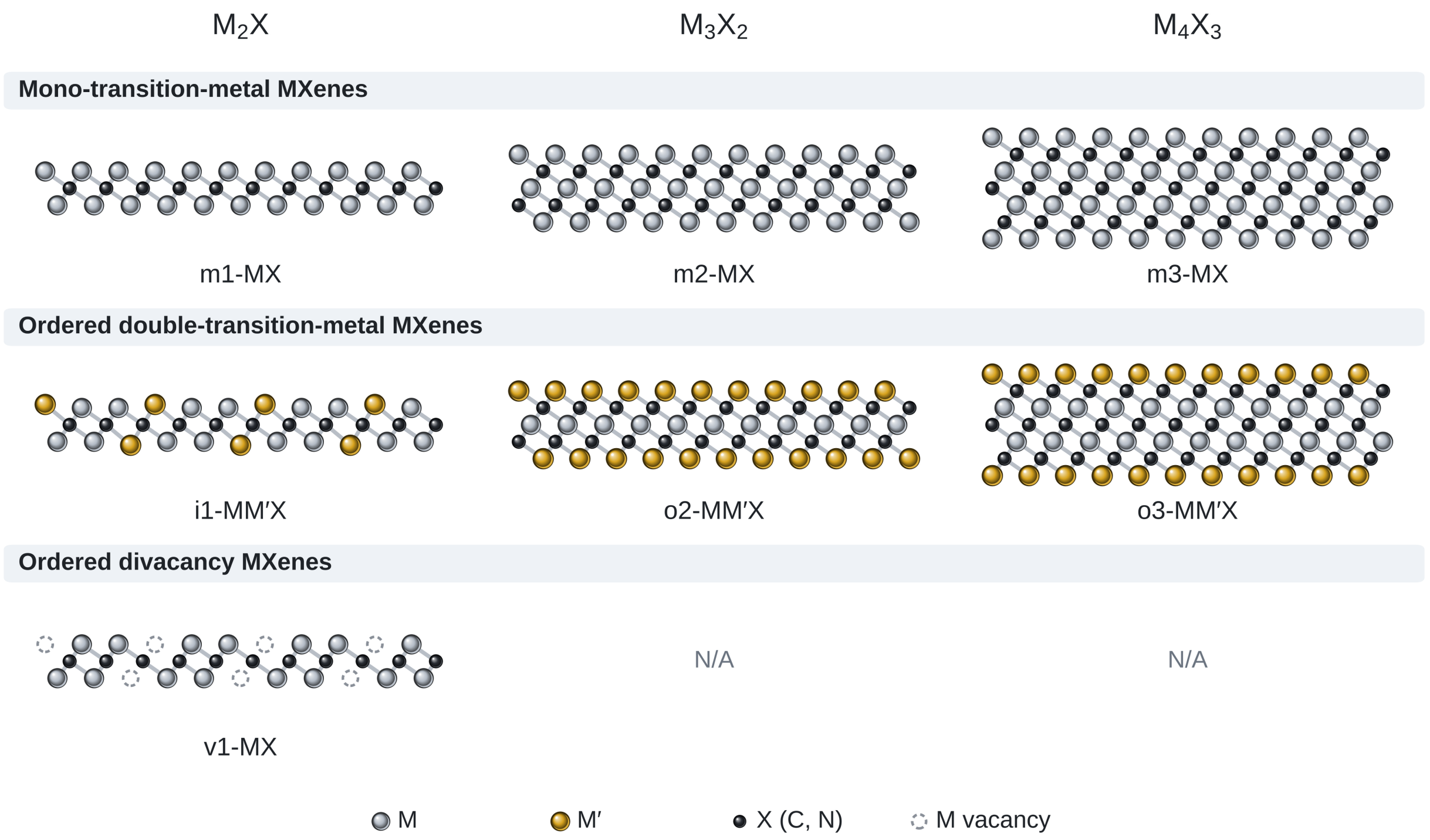

**Figure 1** Schematic illustration of the structural diversity of ordered MXenes. MXenes have the general formula $M_{n+1}X_n$, where M is an early transition metal and X is carbon or nitrogen. The crystal structure varies with n: for n = 1, M atoms adopt hcp stacking; for n > 1, they form an fcc-like arrangement with X occupying octahedral interstitial sites. The ordered MXene family includes mono-transition metal MXenes (m-MXenes), in-plane ordered double-transition metal MXenes (i-MXenes), in-plane ordered metal-divacancy MXenes (v-MXenes), and out-of-plane ordered double-transition-metal MXenes (o-MXenes). The silver, gold, and black spheres represent M, M', and X, respectively. Dashed circles mark ordered metal vacancies**.** To ensure clarity, we adopt a practical naming convention. From left to right: (i) a letter denoting the MXene category; (ii) the value of n; and (iii) element symbols—two for binary MXenes (metal, X), and three for ternary MXenes (two metals, X). For o-MXenes, the first metal represents the inner layer(s) and the second metal the outer layer(s); for i-MXenes, the first and second metals occupy two-thirds and one-third of the in-plane metal sites, respectively.

## Computational details

Meta-GGA spin-polarized DFT calculations were performed using the Vienna Ab-initio Simulation Package (VASP)[20] with the $r^2$SCAN exchange-correlation functional[21] and projector augmented wave (PAW) pseudopotentials[22] under periodic boundary conditions. The electronic wave functions were expanded over a plane wave basis set with an energy cutoff of 680 eV as suggested in `MPScanRelaxSet`.[18] The Brillouin zones were sampled during geometry optimization using a Γ-point centered k-point mesh with a grid density of at least 2000/(atoms/unit cell) to appropriately model the predominantly metallic MXenes. We used the default pseudopotentials and initial magnetic moments specified by MP.[12,18] The relaxation of electronic degrees of freedom was terminated when the change in total energies between two consecutive electronic steps was < $10^{-6}$ eV. The structural relaxations were completed when the residual force on each atom was < 0.01 eV/Å.

After geometry optimization, the layer groups of all MXenes were analyzed using the `SpacegroupAnalyzer`,[18] which determines space groups through symmetry-finding operations. Because `SpacegroupAnalyzer` supports only space group identification, we subsequently mapped the resulting space groups to their corresponding layer groups.[18] Laminated MXenes inherit the hexagonal $P6_3/mmc$ space group symmetry from their parent MAX phases,[2] the ideal delaminated m- and o- MXenes adopt the p-3m1 layer group and v-MXenes and i-MXenes crystallize in the c2/m11 layer group.

Competing phases with $E_{hull} \leq 70$ meV/atom and 200 meV/atom were retrieved from the MP repository to accommodate the thermodynamic metastability of carbides and nitrides, respectively.[12,23] We then recalculated competing phases using the computational approach described above to maintain consistent $r^2$SCAN-based calculations, which are not fully available in MP. The formation and decomposition Gibbs energies and energies above the hull were evaluated using the `PhaseDiagram` module of the pymatgen library.[18] The $N_2$(g) reference chemical potential was corrected using temperature-dependent entropy values from the NIST-JANAF tables to account for gas-phase thermodynamic effects.[24]

To explore non-equilibrium synthesis routes, we assessed the stabilization of MXenes under **(a)** elevated temperature (*t-stabilization*), **(b)** applied pressure (*p-stabilization*), and **(c)** modified chemical potential of X (C or N) (*$\mu_X$-stabilization*). For **(a)**, we used the `GibbsComputedStructureEntry` class[18,25] to generate phase diagram entries, which uses machine-learned entropic contributions to the Gibbs energy over the 300–2000 K temperature range in 100 K increments. In the case of **(b)**, we examined the pressure-dependent stability of MXenes over the 1–10 GPa pressure range in 1 GPa increments, which is experimentally accessible,[26] by applying Amsler's approximation[27] of enthalpy (H) to all entries in a given phase diagram at a specified pressure (p), as shown in Equation 1.

$$H(p) = E_0 + V(p_0)\Delta p \qquad (1)$$

where $p_0$ = 0 atm, $E_0$ is the DFT-calculated total energy at $p_0$, $\Delta p = p - p_0$, and $V(p_0)$ is the DFT-calculated volume per atom at p = 0 atm. This approximation for the enthalpy was applied to all phases in each phase diagram. An exception was made for $N_2$(g), whose chemical potential is derived from its pressure-dependent fugacity using empirical data from Horvath-Bordon et al.[26] Lastly, for **(c)**, we used the `GrandPotentialPhaseDiagram` class,[18,28,29] which constructs phase diagrams open to one or more elements to calculate the grand potential. In this study, we adjusted $\mu_X$ by up to 1 eV/atom, a range that can be experimentally realized by plasma cracking of X containing complex precursors.[17,30]

## Results and discussion

After geometry optimization, all m-, o-, and v-MXenes retained their initial ideal symmetries, as described in the computational details section. In contrast, among the 312 i-MXenes, 284 retained their ideal symmetry, while 13 relaxed to cm11, which still falls within the symmetry domain of layered structures. However, 15 i-MXenes transitioned to p-1 and p1 crystallographic groups, both of which lie outside the domain of layered structures. Notably, all 15 symmetry-broken MXenes and the 13 with cm11 symmetry are i-MXenes with one of the 5d transition metals (Hf, Ta, and W), as shown in **Table S1**, occupying the minority metal (M') site, which accounts for one third of the metal atoms in the i-MXenes. We attribute this to the lanthanide contraction effect, where weak 4f electron shielding contracts the atomic radii, and thus the 5d orbitals. As a result, 5d elements experience a higher effective nuclear charge and lower reactivity, this suggests that lanthanide contraction may contribute to destabilizing the ordered i-MXenes motif.[31] The $r^2$SCAN-computed bandgaps predict that the vast majority of the studied MXenes are semi-metals, as illustrated in **Figure 2(A)**. Only 27 have bandgaps between 0.1 and 0.17 eV, and that just one, v1-ScC, exhibits a notably wider bandgap of 0.47 eV. The near-step empirical cumulative distribution function (ECDF) at ~0 eV indicates that most bare MXenes have negligible bandgaps. This is expected because, without surface terminations, the electronic structure is dominated by transition-metal states that form broad bands in the in-plane metal network. Delamination does not remove that in-plane connectivity, so the density of states near the Fermi level usually persists. In other words, metallicity is the default outcome for unterminated MXenes because the surface metal atoms remain electronically active rather than being passivated.[32] For instance, Sun et al.[33] predict termination-dependent metal–insulator transitions and wide bandgaps in specific terminated double-transition-metal MXenes, whereas the unterminated analogs remain metallic.

While ~60% of the studied MXenes have negative formation energies relative to their elemental references, this does not imply stability with respect to competing phases. All MXenes exhibit positive decomposition Gibbs energies and lie above the convex hull, exceeding accepted metastability thresholds of 70 meV/atom for carbides and 200 meV/atom for nitrides.[12,23] As a result, none of the MXenes considered herein are thermodynamically stable against decomposition into competing phases. This is consistent with prior reports that MXenes are thermodynamically inaccessible and lie above the convex hull under equilibrium conditions.[2–4,34] For benchmarking, we compared the formation energies and $E_{hull}$ values of the 72 available MXene entries in the C2DB database with our corresponding calculated values, as summarized in **Table S2**. The mean absolute errors (MAEs) were 0.15 eV/atom for formation energy and 0.08 eV/atom for $E_{hull}$, which are reasonable MAEs given the differences in exchange-correlation functionals; PBE in C2DB and $r^2$SCAN in our work. Notably, neither our calculations nor those from the C2DB database yielded negative/zero $E_{hull}$ values, confirming that the studied MXenes are thermodynamically metastable under equilibrium conditions, consistent with prior observations for many 2D materials.[13,14]

Our study of the temperature-dependent stability of MXenes revealed that 115 MXenes become thermodynamically stable at elevated temperatures, where all competing phases in the diagram were treated consistently within `GibbsComputedStructureEntry`. Most of the 115 t-stabilized MXenes are carbides, as tabulated in **Table S3**. This highlights the composition-dependent stabilizing effect of entropic contributions. If a given MXene and its competing phases are close in energy at 0 K, even moderate differences in their entropy will reorder their stability and potentially change which configuration is lowest in Gibbs free energy. Thus, the large number of t-stabilized MXenes suggests that a significant fraction of metastable MXenes may be pulled onto the convex hull when temperature-dependent terms are included. The t-stabilized MXenes are primarily m- and o-MXenes with only one i-MXene, i1-YScC, as presented in **Figure 2(B)**. Notably, all m-MXenes (i.e., binary MXenes that can exist in both binary and ternary MXene chemical spaces) exhibited the same onset temperature, indicating that the stabilization temperature is agnostic to the chemical space within the resolutions allowed by the 100 K increments of `GibbsComputedStructureEntry`. For t-stabilized MXene carbides and nitrides with the same n and metal species arrangement, the nitrides generally require higher onset temperatures, as tabulated in **Table S3**. This trend holds true, with the exception of the o-MXenes with n=3, M=Cr, and M'=Mn. In some cases where the carbides and nitrides exhibited the same onset temperatures. This is likely due to the

`GibbsComputedStructureEntry` class, which only allows 100 K temperature increments, which likely cannot resolve the actual onset temperatures. For MXenes with identical composition and metal-site configuration, lower n values typically correspond to higher onset temperatures. This trend does not hold for the o-MXene carbides with M ∈ {Sc, Nb, Ta} and M'=Mn.

For the p-stabilized MXenes, a total of 23 were predicted to be thermodynamically stabilized at 10 GPa. These were exclusively m- and o-MXenes and mostly carbides, as tabulated in **Table S3**. Furthermore, 20 are common with the temperature-stabilizable MXenes, as shown in **Figure 2(B)**. Pressure stabilizes MXenes whose per-atom volume is sufficiently small relative to competing phases such that the $V\Delta p$ contributions lower their H(p) relative to competing phases.[27] The large overlap between t- and p-stabilized MXenes indicates that both T and p can be used to push those MXenes onto the convex hull, though via either their $-s\Delta T$ or $V\Delta p$ contributions. The trends in onset pressures closely mirror those observed for t-stabilized MXenes: carbides generally exhibit lower onset pressures than their nitride counterparts, and, for a given MXene composition and configuration, MXenes with lower n tend to require higher stabilizing pressures. An exception to this trend is observed for the o-MXene carbides with M=Nb and M'=Sc.

As for the $\mu_X$-stabilized MXenes, the Mn–Fe–C chemical space was the only one in which stabilized MXenes (with n = 3) were predicted. Specifically, only m-MXene carbides with M ∈ {Mn, Fe}, and the o3-MnFeC MXene. According to our calculations, the suitable window of $\mu_C$ for the coexistence of these three MXenes ranges from –6 to 1 eV/C atom, with 1 eV/C atom representing an upper bound consistent with experimentally accessible carbon activities (e.g., plasma-assisted or activated precursor environments), beyond which conditions become increasingly impractical for materials synthesis. Chemical-potential stabilization can only occur when there is a finite range of $\mu_X$ over which the target phase is not outcompeted by either X-rich or X-poor competing phases. For the MXenes studied here, the optimum $\mu_X$ window is generally absent because competing phases with higher X stoichiometry shift more strongly in grand potential with increasing $\mu_X$, making them thermodynamically favorable over the target phase with only modest changes in $\mu_X$. The fact that $\mu_X$-stabilization is predicted for only a small set of Mn–Fe–C MXenes is therefore consistent with a narrow-chemistry scenario in which the target MXene lies sufficiently close in free energy and is more sensitive to $\mu_X$ changes than the competing phases.

Among the 121 stabilized MXenes, 90 were carbides and 31 were nitrides, a trend consistent with experimental observations that report a higher prevalence of carbide MXenes over nitrides,[2,35] as shown in **Figure 2(B)**. This is consistent with the reported challenge of synthesizing nitrides, rooted in the exceptional stability of $N_2$ provided by its triple bond and the high decomposition temperatures of solid-state nitride precursors.[17] In this study, the stabilized MXenes have 9 and 25 experimentally and computationally reported cases in the literature,[10] respectively, as summarized in **Table S3**. Given that m- and o-MXenes are the dominant stabilizable categories, **Figure 2(C)** and **Figure 2(D)** present the stabilizable compositions of m-MXenes as a binary map and o-MXenes as a heat map, respectively.

**Figure 2(C)** shows that Sc and Y are the most frequently present elements among the stabilizable m-MXenes. This may be because Sc and Y are early transition metals (group 3), which are relatively electropositive, and often form stable carbide/nitride frameworks. In a layered structure, such metals can better accommodate the MXene motif without incurring as large of a free energy penalty relative to competing phases, placing these MXenes closer to the near-stable regime where temperature or pressure can bring these phases onto the convex hull.[34,35] In **Figure 2(C),** the n=3 stoichiometry is the most frequently represented among the stabilizable m-MXene carbides and nitrides. This trend aligns with our earlier observation that higher n values generally correspond to lower onset temperatures, rendering such MXenes more thermodynamically accessible. Increasing n raises the fraction of interior metal atoms relative to surface atoms. Delamination and the absence of terminating groups impose the largest energetic penalties by increasing the fraction of undercoordinated metal atoms with dangling bonds. As n increases, the energetic penalty of forming the two surfaces is diluted per atom, pushing the free energy of the structure closer to the convex hull, where other bulk-like layered phases reside. This makes stabilization by temperature or pressure more feasible.[35,36] In contrast, V, Mo, and W do not form any stabilizable m-MXenes under the studied conditions, which implies that V/Mo/W chemistries have sufficiently stable competing phases in their respective chemical spaces such that the MXene lies too far above the convex hull for the explored levers of t, p, and $\mu_X$ to stabilize it relative to the competing phases.

**Figure 2(D)** shows that Zr and Hf are the most common inner-layer metals in stabilizable MXene carbides and nitrides. The inner layers define the backbone of the MXene as they set the equilibrium lattice parameters and most of the metal–X framework energetics. Group-4 metals (Zr/Hf) are classic carbide/nitride framework formers and tend to stabilize metal–X networks. In an o-MXene, placing a strong framework-forming metal into an inner layer reduces the energetic penalty of maintaining a layered motif relative to competing phases.[34,35] Cr and Mn are the most frequent outer-layer metals associated with stabilizable MXene carbides and nitrides. Surface atoms primarily determine the scale of the free energy changes resulting from surface formation and reconstruction. Cr/Mn (middle transition metals) can accommodate a wide spectrum of bonding and oxidation environments and, in practice, often stabilize the coordination environments at and near surfaces more than other transition metals.[35,36] Generally, Fe and W appear least frequently in stabilizable MXenes under the studied conditions. Fe, as a later 3d transition metal, often results in Fe-containing MXenes that compete with numerous low-energy phases. Moreover, magnetic and electronic contributions can preferentially stabilize alternative competing phases within the Fe-containing chemical space, making it more difficult to stabilize an Fe-containing MXene. W, a 5d element, also appears in the symmetry-breaking list (**Table S1**) for several ordered i-MXenes, consistent with the broader observation from the results reported in **Table S1** that heavy-element substitutions in ordered layered structures can render the free energies of these phases structurally and thermodynamically sensitive. Importantly, this should be viewed as a probabilistic tendency rather than a strict rule. [34] Further, MXenes with c2/m11 layer symmetry, such as i- and v-MXenes, which involve in-plane metal substitution or vacancies, are more difficult to stabilize compared to p-3m1 MXenes with pristine metal layers. Finally, we observed that all stabilizable MXene nitrides in this study have corresponding carbide counterparts, regardless of the n value, that are also stabilizable. However, the opposite is not necessarily true; many stabilizable carbides do not have stabilizable nitride counterparts.

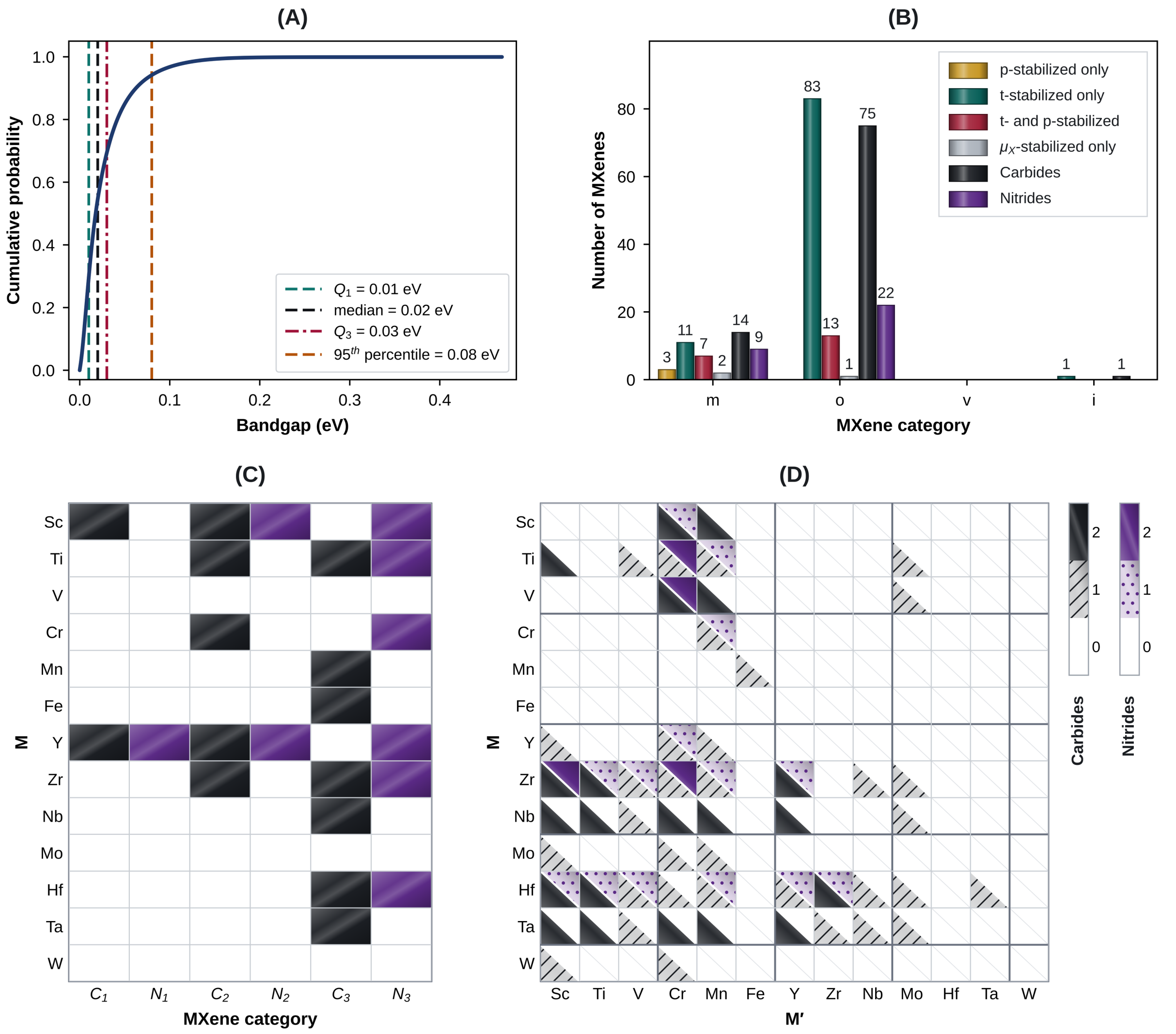

**Figure 2: (A)** Cumulative distribution of the r$^2$SCAN-calculated bandgaps for the studied MXenes, shown as a kernel-smoothed curve; dashed lines mark $Q_1$ and $Q_3$ (the first and third quartiles), the median and the 95th percentile, all computed from the raw data. The vast majority exhibit semi-metallic character, while only 28 MXenes exhibit bandgaps above 0.1 eV (27 between 0.1 and 0.172 eV, and one at 0.47 eV). **(B)** t-, p-, and $\mu_X$-dependent stabilization of the 121 stabilizable MXenes. The 115 t-stabilized MXenes include m- and o-categories, and only one i-MXene, i1-YScC. p-stabilization at 10 GPa revealed 23 stable MXenes, with 20 MXenes common with the t-stabilizable MXenes. Additionally, only three $\mu_X$-stabilized MXenes were identified in the Mn–Fe–C chemical space. Overall, 90 of the stabilizable MXenes are carbides and 31 are nitrides, consistent with the greater experimental accessibility of carbide MXenes. **(C)** Binary map of stabilizable m-MXene compositions, where the figure highlights the metal atoms and n values for stabilizable m-MXene carbides (graphite) and nitrides (purple). **(D)** Heatmap of stabilizable o-MXene compositions. The map shows combinations of M (y-axis) and M' (x-axis) metals for stabilizable carbides and nitrides. In each cell, the bottom-left triangle indicates the presence of one (hatched grey) or two (solid graphite) stabilizable o-MXene carbides, while the top-right triangle indicates one (dotted lavender) or two (solid purple) stabilizable o-MXene nitrides. Zr and Hf dominate M (inner layers), while Cr and Mn are most common in M' (outer layers); Fe and W appear least frequently overall (Fe never occurs as M, and W never as M'). All stabilizable nitrides have carbide counterparts, but not all stabilizable carbides have nitride counterparts.

## Conclusions

MXenes hold immense promise for diverse applications. However, their synthesis remains a challenge due to their broad compositional and structural space and the limited understanding of what governs their stability, and experimental efforts still rely heavily on chemical intuition and heuristics. In this study, we employed DFT with the $r^2$SCAN functional to generate a comprehensive dataset of 1,040 ordered binary and ternary bare delaminated MXenes, encompassing m-MXenes, i-MXenes, v-MXenes, and o-MXenes. Our analysis provides detailed thermodynamic data, including formation and decomposition Gibbs energies. All of the MXenes lie above the hull under ambient conditions within the $r^2$SCAN thermodynamic framework, indicating that they are metastable under standard thermodynamic conditions. However, we identified 115 MXenes, mostly carbides, that are stabilizable at elevated temperatures (500–2000 K), with m- and o-MXenes dominating this group. Additionally, 23 MXenes, also primarily carbides, are stabilized under a pressure of 10 GPa, while specific Mn–Fe–C ternary MXenes are stabilized by adjusting the chemical potential of carbon within an experimentally feasible range of –6 to 1 eV/atom. These results highlight the potential of non-equilibrium synthesis routes, such as high-temperature or high-pressure techniques, to unlock the synthesis of new MXenes.

This study offers actionable guidance for experimentalists, suggesting conditions such as high-temperature processing for carbides or pressure-based methods to stabilize select compositions. Furthermore, the dataset serves as a valuable resource for the MXene community, enabling the development of machine learning models to predict stability, thus reducing the reliance on heuristic approaches.

As an outlook, this dataset can be expanded to include disordered and terminated MXenes, which significantly influence stability and properties. Experimental validation of the predicted stable MXenes is also necessary to confirm these findings and guide synthesis. In summary, this study provides a high-fidelity computational dataset and identifies potentially synthesizable MXene candidates that are stabilized by non-equilibrium conditions. This dataset provides a foundation for future autonomous experimental workflows for the systematic exploration of MXenes for next-generation technologies.

## Data Records

The dataset is openly available in Zenodo under a CC BY 4.0 license (https://doi.org/10.5281/zenodo.24116692). It contains a table of all 1,040 MXenes in CSV and JSON formats, the DFT-optimized structure of each MXene as a VASP CONTCAR file, a README describing each column, and SHA-256 checksums for all files. Each of the 1,040 entries has a unique identifier consisting of two parts separated by a dash. The first part is a letter indicating the MXene category, as described in the Introduction, followed by the value of n in the corresponding category formula. For binary (m- and v-) MXenes, the second part gives the M and X elements, while for ternary (o- and i-) MXenes it gives the M, M', and X elements, respectively. For o-MXenes, M denotes the inner metal layer(s) and M' the outer metal layers; for i-MXenes, M occupies two-thirds of the metal sites and M' the remaining one-third. For example, the o-MXene with M = Sc, M' = Y, X = C and n = 2 is “o2-ScYC.” The relaxed structure of each MXene is stored in a folder named by this identifier. For each entry, the table lists $\Delta G_f$ (eV/atom), $\Delta G_d$ (eV/atom), $E_{hull}$ (eV/atom), the decomposition products, the band gap (eV), and the layer group. $E_{hull}$, $\Delta G_f$, and $\Delta G_d$ are the energy above the hull and the formation and decomposition Gibbs energies of the DFT-optimized structures, computed relative to all competing phases, the elemental references, and all competing phases, respectively. The decomposition products are the competing phases that dominate under equilibrium conditions and inhibit formation of the corresponding MXene. The minimum stabilizing temperatures and pressures of the t- and p-stabilized MXenes are listed in Table S3 of the Supporting Information.

## CONFLICT OF INTEREST

The authors declare no conflicts of interest.

## ASSOCIATED CONTENT

**Supporting Information:**
Layer-group symmetry breaking of the relaxed MXenes (Table S1); benchmarking of $\Delta G_f$ and $E_{hull}$ against C2DB (Table S2); t- and p-stabilized MXenes with their minimum stabilizing temperatures and pressures (Table S3).

**Data Availability:**
The dataset supporting this study is available in Zenodo at https://doi.org/10.5281/zenodo.24116692.

## ACKNOWLEDGMENTS

We gratefully acknowledge the support from the National Science Foundation, Award No. NSF CBET-2323065. The views expressed in this article do not necessarily represent the views of the National Science Foundation nor the U.S. Government. This work used Alpine at the University of Colorado Boulder Research Computing's third-generation high performance computing (HPC) cluster through allocation ucb404_peak2 and the NSF Anvil CPU at Purdue University through allocation MAT240100 from the Advanced Cyberinfrastructure Coordination Ecosystem: Services & Support (ACCESS) program, which is supported by NSF grants 2138259, 2138286, 2138307, 2137603, and 2138296.

# Supplemental Information

# High-Fidelity In Silico Dataset of Ordered Binary and Ternary MXenes with Predictive Synthesis Guidance

Ahmed H. Biby[1] and Charles B. Musgrave* [2,3,4]
[1]Materials Science and Engineering Program, University of Colorado Boulder, Boulder, CO 80309, USA
[2]Department of Chemical and Biological Engineering, University of Colorado, Boulder, Colorado 80303, USA
[3]Department of Chemical Engineering, University of Utah, Salt Lake City, Utah, 84112, USA
[4]Department of Materials Science and Engineering, University of Utah, Salt Lake City, Utah, 84112, USA

**Table S1:** MXenes that exhibit symmetry breaking and deviate from their initial ideal layer group symmetry.

| MXene | layer group |
|---|---|
| **i1-FeTaC** | p-1 |
| **i1-FeTaN** | p1 |
| **i1-FeWN** | p-1 |
| **i1-MnHfC** | p-1 |
| **i1-MnWN** | p-1 |
| **i1-MoWN** | p-1 |
| **i1-NbTaC** | p-1 |
| **i1-NbTaN** | p-1 |
| **i1-NbWN** | p-1 |
| **i1-TiHfN** | p-1 |
| **i1-TiTaN** | p-1 |
| **i1-ZrHfN** | p-1 |
| **i1-ZrTaN** | p-1 |
| **i1-ZrWC** | p-1 |
| **i1-ZrWN** | p-1 |

**Table S2:** Benchmarking of $\Delta G_f$ and $E_{hull}$ values for 72 MXenes available in the C2DB database, compared against our corresponding $r^2$SCAN-based calculations.

| MXene | $\Delta G_f$ (eV/atom) of this study | $\Delta G_f$ (eV/atom) of C2DB | $E_{hull}$ (eV/atom) of this study | $E_{hull}$ (eV/atom) of C2DB |
|---|---|---|---|---|
| **m1-CrC** | 0.56 | 0.79 | 0.67 | 0.88 |
| **m1-CrN** | 0.40 | 0.53 | 0.99 | 0.89 |
| **m1-HfC** | -0.09 | 0.01 | 0.66 | 0.62 |
| **m1-HfN** | -0.58 | -0.44 | 0.83 | 0.69 |
| **m1-MnC** | 0.49 | 0.70 | 0.53 | 0.76 |
| **m1-MnN** | 0.18 | 0.59 | 0.63 | 0.82 |
| **m1-MoC** | 0.87 | 0.79 | 1.01 | 0.97 |
| **m1-MoN** | 0.69 | 0.66 | 1.05 | 0.99 |
| **m1-NbC** | 0.36 | 0.28 | 0.90 | 0.84 |
| **m1-NbN** | 0.12 | 0.03 | 1.11 | 1.03 |
| **m1-ScC** | -0.12 | -0.06 | 0.40 | 0.38 |
| **m1-ScN** | -0.91 | -0.81 | 0.55 | 0.49 |
| **m1-TaC** | 0.33 | 0.32 | 1.01 | 0.96 |
| **m1-TaN** | 0.18 | 0.17 | 1.20 | 1.14 |
| **m1-TiN** | -0.67 | -0.59 | 0.78 | 0.74 |
| **m1-VC** | 0.30 | 0.34 | 0.84 | 0.79 |
| **m1-VN** | 0.04 | -0.06 | 1.12 | 0.89 |
| **m1-WC** | 1.08 | 0.96 | 1.19 | 1.09 |
| **m1-WN** | 0.95 | 0.87 | 1.26 | 1.06 |
| **m1-YC** | -0.02 | 0.04 | 0.40 | 0.37 |
| **m1-YN** | -0.80 | -0.70 | 0.54 | 0.46 |
| **m1-ZrN** | -0.64 | -0.54 | 0.78 | 0.69 |
| **m2-CrC** | 0.21 | 0.54 | 0.33 | 0.62 |
| **m2-CrN** | -0.23 | 0.21 | 0.47 | 0.61 |
| **m2-HfC** | -0.53 | -0.37 | 0.37 | 0.36 |
| **m2-HfN** | -1.12 | -0.88 | 0.56 | 0.48 |
| **m2-MnC** | 0.28 | 0.44 | 0.31 | 0.50 |
| **m2-MnN** | -0.15 | 0.18 | 0.37 | 0.39 |
| **m2-MoC** | 0.52 | 0.46 | 0.67 | 0.63 |
| **m2-MoN** | 0.33 | 0.32 | 0.77 | 0.71 |
| **m2-NbC** | -0.02 | -0.03 | 0.56 | 0.58 |
| **m2-NbN** | -0.32 | -0.40 | 0.78 | 0.69 |
| **m2-ScC** | -0.39 | -0.28 | 0.20 | 0.21 |
| **m2-ScN** | -1.45 | -1.24 | 0.31 | 0.31 |
| **m2-TaC** | -0.05 | 0.01 | 0.62 | 0.64 |
| **m2-TaN** | -0.28 | -0.31 | 0.89 | 0.77 |
| **m2-TiC** | -0.50 | -0.38 | 0.35 | 0.36 |
| **m2-TiN** | -1.18 | -0.99 | 0.47 | 0.49 |
| **m2-VC** | -0.09 | 0.02 | 0.50 | 0.52 |
| **m2-VN** | -0.56 | -0.37 | 0.63 | 0.66 |
| **m2-WC** | 0.78 | 0.59 | 0.91 | 0.75 |
| **m2-WN** | 0.59 | 0.55 | 0.97 | 0.77 |
| **m2-YC** | -0.25 | -0.14 | 0.19 | 0.20 |
| **m2-YN** | -1.29 | -1.09 | 0.31 | 0.29 |
| **m2-ZrC** | -0.53 | -0.42 | 0.31 | 0.28 |
| **m2-ZrN** | -1.16 | -0.97 | 0.48 | 0.51 |
| **m3-CrC** | 0.19 | 0.37 | 0.30 | 0.45 |
| **m3-CrN** | -0.42 | -0.12 | 0.34 | 0.29 |
| **m3-FeC** | 0.74 | 0.73 | 0.74 | 0.73 |
| **m3-FeN** | 0.34 | 0.21 | 0.39 | 0.27 |
| **m3-HfC** | -0.69 | -0.53 | 0.27 | 0.25 |
| **m3-HfN** | -1.40 | -1.19 | 0.37 | 0.27 |
| **m3-MnC** | 0.22 | 0.39 | 0.25 | 0.45 |
| **m3-MnN** | -0.27 | 0.00 | 0.28 | 0.19 |
| **m3-MoC** | 0.41 | 0.35 | 0.57 | 0.52 |
| **m3-MoN** | 0.17 | -0.12 | 0.63 | 0.30 |
| **m3-NbC** | -0.22 | -0.27 | 0.39 | 0.37 |
| **m3-NbN** | -0.49 | 0.57 | 0.65 | 0.56 |
| **m3-ScC** | -0.41 | -0.32 | 0.22 | 0.19 |
| **m3-ScN** | -1.66 | -1.49 | 0.22 | 0.18 |
| **m3-TaC** | -0.28 | -0.26 | 0.39 | 0.36 |
| **m3-TaN** | -0.45 | -0.42 | 0.77 | 0.70 |
| **m3-TiC** | -0.63 | -0.52 | 0.25 | 0.25 |
| **m3-TiN** | -1.42 | -1.26 | 0.32 | 0.15 |
| **m3-VC** | -0.22 | -0.13 | 0.40 | 0.38 |
| **m3-VN** | -0.75 | -0.59 | 0.48 | 0.47 |
| **m3-WC** | 0.67 | 0.55 | 0.80 | 0.72 |
| **m3-WN** | 0.43 | 0.63 | 0.84 | 0.87 |
| **m3-YC** | -0.25 | -0.16 | 0.20 | 0.19 |
| **m3-YN** | -1.50 | -1.30 | 0.22 | 0.19 |
| **m3-ZrC** | -0.66 | -0.55 | 0.24 | 0.20 |
| **m3-ZrN** | -1.42 | -1.28 | 0.32 | 0.29 |

**Table S3:** t-and p-stabilized MXenes, listing the corresponding minimum stabilizing temperatures (K) and pressures (GPa) together with available experimental and computational reports in ref. 1.

| MXene | Minimum stabilizing temperature (K) | Minimum stabilizing pressure (GPa) | Experimentally reported | Computationally reported |
|---|---|---|---|---|
| **i1-YScC** | 1900 | - | - | - |
| **m1-ScC** | 1900 | 8 | - | ✓ |
| **m1-YC** | 1800 | 6 | - | - |
| **m1-YN** | - | 8 | - | - |
| **m2-CrC** | 1200 | - | - | - |
| **m2-ScC** | 800 | 7 | - | - |
| **m2-ScN** | - | 10 | - | - |
| **m2-TiC** | 1800 | - | ✓ | ✓ |
| **m2-YC** | 800 | 4 | - | - |
| **m2-YN** | - | 8 | - | - |
| **m2-ZrC** | 1800 | 10 | ✓ | ✓ |
| **m3-CrN** | 1900 | - | - | - |
| **m3-HfC** | 1500 | - | - | - |
| **m3-HfN** | 2000 | - | - | - |
| **m3-NbC** | 2000 | - | ✓ | - |
| **m3-ScN** | 1100 | 10 | - | - |
| **m3-TaC** | 2000 | - | ✓ | - |
| **m3-TiC** | 1200 | - | - | - |
| **m3-TiN** | 1700 | - | ✓ | - |
| **m3-YN** | 1100 | 7 | - | - |
| **m3-ZrC** | 1200 | - | - | - |
| **m3-ZrN** | 1700 | - | - | - |
| **o2-HfScC** | 1400 | 7 | - | - |
| **o2-HfTiC** | 1700 | - | - | - |
| **o2-HfZrC** | 1800 | 10 | - | - |
| **o2-MoCrC** | 1200 | - | - | - |
| **o2-MoMnC** | 1800 | - | - | - |
| **o2-MoScC** | 1300 | 10 | - | - |
| **o2-NbCrC** | 1200 | - | - | ✓ |
| **o2-NbMnC** | 1500 | - | - | - |
| **o2-NbScC** | 800 | 7 | - | - |
| **o2-NbTiC** | 1900 | - | - | ✓ |
| **o2-NbYC** | 1500 | 8 | - | - |
| **o2-ScCrC** | 800 | - | - | - |
| **o2-ScMnC** | 1000 | - | - | - |
| **o2-TaCrC** | 1400 | - | - | ✓ |
| **o2-TaMnC** | 1400 | - | - | - |
| **o2-TaScC** | 800 | 7 | - | - |
| **o2-TaTiC** | 1800 | - | - | ✓ |
| **o2-TaYC** | 1200 | 7 | - | - |
| **o2-TiCrN** | 2000 | - | - | - |
| **o2-TiScC** | 1800 | - | - | - |
| **o2-VCrC** | 1400 | - | ✓ | ✓ |
| **o2-VCrN** | 1700 | - | - | - |
| **o2-VMnC** | 1300 | - | - | - |
| **o2-WCrC** | 1700 | - | - | - |
| **o2-WScC** | 1300 | 10 | - | - |
| **o2-YCrC** | 1200 | - | - | - |
| **o2-YMnC** | 2000 | - | - | - |
| **o2-YScC** | 1100 | - | - | - |
| **o2-ZrCrN** | 2000 | - | - | - |
| **o2-ZrScC** | 1400 | 7 | - | - |
| **o2-ZrScN** | 1800 | - | - | - |
| **o2-ZrTiC** | 1900 | - | - | - |
| **o2-ZrYC** | 1800 | - | - | - |
| **o3-CrMnC** | 1300 | - | - | - |
| **o3-CrMnN** | 1100 | - | - | - |
| **o3-HfCrC** | 900 | - | - | - |
| **o3-HfMnC** | 1400 | - | - | - |
| **o3-HfMnN** | 1800 | - | - | - |
| **o3-HfMoC** | 1900 | - | - | ✓ |
| **o3-HfNbC** | 1700 | - | - | - |
| **o3-HfScC** | 700 | 6 | - | - |
| **o3-HfScN** | 1400 | - | - | - |
| **o3-HfTaC** | 1900 | - | - | - |
| **o3-HfTiC** | 1300 | - | - | - |
| **o3-HfTiN** | 1700 | - | - | - |
| **o3-HfVC** | 1800 | - | - | - |
| **o3-HfVN** | 2000 | - | - | - |
| **o3-HfYC** | 1000 | - | - | - |
| **o3-HfYN** | 1400 | - | - | - |
| **o3-HfZrC** | 1100 | - | - | - |
| **o3-HfZrN** | 1500 | - | - | - |
| **o3-NbCrC** | 1000 | - | - | ✓ |
| **o3-NbMnC** | 1600 | - | - | - |
| **o3-NbMoC** | 1800 | - | ✓ | ✓ |
| **o3-NbScC** | 700 | 8 | - | - |
| **o3-NbTiC** | 1800 | - | - | ✓ |
| **o3-NbVC** | 1900 | - | - | ✓ |
| **o3-NbYC** | 1200 | - | - | - |
| **o3-ScCrC** | 800 | - | - | - |
| **o3-ScCrN** | 1400 | - | - | - |
| **o3-ScMnC** | 1100 | - | - | - |
| **o3-TaCrC** | 1000 | - | - | ✓ |
| **o3-TaMnC** | 1500 | - | - | - |
| **o3-TaMoC** | 1600 | - | - | ✓ |
| **o3-TaNbC** | 1600 | - | - | ✓ |
| **o3-TaScC** | 500 | 6 | - | - |
| **o3-TaTiC** | 1600 | - | - | ✓ |
| **o3-TaVC** | 1900 | - | - | ✓ |
| **o3-TaYC** | 800 | - | - | - |
| **o3-TaZrC** | 1700 | - | - | - |
| **o3-TiCrC** | 900 | - | - | ✓ |
| **o3-TiCrN** | 1600 | - | - | - |
| **o3-TiMnC** | 1000 | - | - | ✓ |
| **o3-TiMnN** | 1200 | - | - | - |
| **o3-TiMoC** | 1600 | - | ✓ | ✓ |
| **o3-TiScC** | 1000 | - | - | - |
| **o3-TiVC** | 1800 | - | - | ✓ |
| **o3-VCrC** | 1300 | - | - | ✓ |
| **o3-VCrN** | 1700 | - | - | - |
| **o3-VMnC** | 1200 | - | - | - |
| **o3-VMoC** | 2000 | - | ✓ | ✓ |
| **o3-YCrN** | 1700 | - | - | - |
| **o3-ZrCrC** | 900 | - | - | - |
| **o3-ZrCrN** | 1700 | - | - | - |
| **o3-ZrMnC** | 1300 | - | - | - |
| **o3-ZrMnN** | 1400 | - | - | - |
| **o3-ZrMoC** | 2000 | - | - | ✓ |
| **o3-ZrNbC** | 1800 | - | - | - |
| **o3-ZrScC** | 700 | 6 | - | - |
| **o3-ZrScN** | 1400 | - | - | - |
| **o3-ZrTiC** | 1300 | - | - | - |
| **o3-ZrTiN** | 1800 | - | - | - |
| **o3-ZrVC** | 1900 | - | - | - |
| **o3-ZrVN** | 2000 | - | - | - |
| **o3-ZrYC** | 900 | - | - | - |
| **o3-ZrYN** | 1400 | - | - | - |